\documentclass[8.5pt,twoside,twocolumn]{article}
\usepackage[super,sort&compress,comma]{natbib}
\usepackage{mhchem,color}
\usepackage{times,mathptmx}
\usepackage{sectsty}
\usepackage{balance}

\usepackage{graphicx} 
\usepackage{lastpage}
\usepackage[format=plain,justification=raggedright,singlelinecheck=false,font=small,labelfont=bf,labelsep=space]{caption}
\usepackage{fancyhdr}
\begin{document}

\thispagestyle{plain}
\fancypagestyle{plain}{
\fancyhead[L]{\includegraphics[height=8pt]{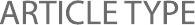}}
\fancyhead[C]{\hspace{-1cm}\includegraphics[height=20pt]{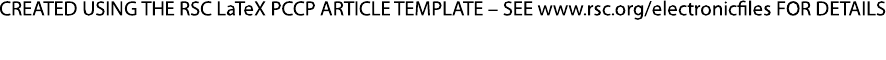}}
\fancyhead[R]{\includegraphics[height=10pt]{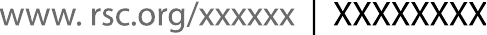}\vspace{-0.2cm}}
\renewcommand{\headrulewidth}{1pt}}
\renewcommand{\thefootnote}{\fnsymbol{footnote}}
\renewcommand\footnoterule{\vspace*{1pt}%
\hrule width 3.4in height 0.4pt \vspace*{5pt}}
\setcounter{secnumdepth}{5}

\makeatletter
\def\subsubsection{\@startsection{subsubsection}{3}{10pt}{-1.25ex plus -1ex minus -.1ex}{0ex plus 0ex}{\normalsize\bf}}
\def\paragraph{\@startsection{paragraph}{4}{10pt}{-1.25ex plus -1ex minus -.1ex}{0ex plus 0ex}{\normalsize\textit}}
\renewcommand\@biblabel[1]{#1}
\renewcommand\@makefntext[1]%
{\noindent\makebox[0pt][r]{\@thefnmark\,}#1}
\makeatother
\renewcommand{\figurename}{\small{Fig.}~}
\sectionfont{\large}
\subsectionfont{\normalsize}

\fancyfoot{}
\fancyfoot[LO,RE]{\vspace{-7pt}\includegraphics[height=9pt]{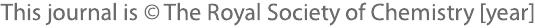}}
\fancyfoot[CO]{\vspace{-7.2pt}\hspace{12.2cm}\includegraphics{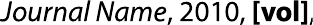}}
\fancyfoot[CE]{\vspace{-7.5pt}\hspace{-13.5cm}\includegraphics{headers/RF.pdf}}
\fancyfoot[RO]{\footnotesize{\sffamily{1--\pageref{LastPage} ~\textbar  \hspace{2pt}\thepage}}}
\fancyfoot[LE]{\footnotesize{\sffamily{\thepage~\textbar\hspace{3.45cm} 1--\pageref{LastPage}}}}
\fancyhead{}
\renewcommand{\headrulewidth}{1pt}
\renewcommand{\footrulewidth}{1pt}
\setlength{\arrayrulewidth}{1pt}
\setlength{\columnsep}{6.5mm}
\setlength\bibsep{1pt}

\twocolumn[
  \begin{@twocolumnfalse}
\noindent\LARGE{\textbf{Time-resolved analysis of strong-field induced plasmon oscillations in metal clusters by spectral interferometry with few-cycle laser fields}}
\vspace{0.6cm}

\noindent\large{\textbf{J\"org K\"ohn and Thomas Fennel$^{\ast}$ }}
\vspace{0.5cm}
\noindent\textit{\small{\textbf{Received Xth XXXXXXXXXX 20XX, Accepted Xth XXXXXXXXX 20XX\newline
First published on the web Xth XXXXXXXXXX 200X}}}

\noindent \textbf{\small{DOI: 10.1039/b000000x}}
\vspace{0.6cm}

\noindent \normalsize{
We propose a scheme for ultrafast real-time imaging of laser-induced collective electron oscillations (Mie plasmons) in gas phase metal clusters by interferometrically stable scanning of two intense few-cycle optical laser pulses. The feasibility of our nonlinear spectral interferometry method with experimentally accessible observables is tested in a theoretical case study on simple-metal clusters (Na$_{147}$). The results show that the plasmon period and lifetime as well as the phase and relative amplitude of the collective electron motion can be extracted with sub-fs resolution. The access to nonlinear response effects, as the demonstrated increase of the plasmon lifetime with laser intensity due to ionization-induced contraction of the electron cloud, opens up vast opportunities for interrogating ultrafast many-particle dynamics in nanosystems under strong laser fields with unprecedented resolution.}
\vspace{0.5cm}
 \end{@twocolumnfalse}
  ]

\section{Introduction}

\footnotetext{\textit{Institut f\"ur Physik, Universit\"at Rostock, Universit\"atsplatz 3, 18051 Rostock, Germany. E-mail: thomas.fennel@uni-rostock.de}}

Coherent control of ultrafast processes in atoms and molecules has become a vital research branch of present-day laser-matter science~\cite{RabSci00,Ric00,NuePCCP07,Sha03,JudPRL92,DanSci03}. Nowadays, the availability of phase-controlled few-cycle pulses and shaped light fields with precisely adjustable amplitude, phase, and polarization~\cite{KraRMP09, BriPRL04,WeiRSI00,KelN03} offers a versatile toolbox for probing and steering electronic dynamics down to sub-fs time scales. Such tailored light fields have been applied for illuminating various aspects of ultrafast nonlinear laser-matter interactions such as quantum path control for generating selected high harmonics \cite{WinRMP08,PfeAPB05} or single attosecond pulses~\cite{SanSci06,ZenPRL07}, tunneling of electrons in atomic photoionization~\cite{UibNat07}, valence electron motion in atoms~\cite{GouNat10}, electron localization in molecular dissociation~\cite{KliSci06}, and selective population of dressed electronic states~\cite{WolAPB06}.

So far, control schemes with sub-fs resolution have been applied predominantly to processes, whose main physics can be understood in an effective single active electron picture. A promising and complementary future direction of ultrafast laser science is the manipulation and imaging of many-particle electron dynamics in the nonlinear response domain. From such efforts, deeper insights into the nature of collective electronic processes in ultrashort laser fields~\cite{DesPRB10,SkoPRL10,MatJCP10}, their ultrafast control~\cite{NguPRA04}, and potential new applications can be expected~\cite{FenRMP10}.

An instructive example for such many-particle dynamics are collective electron oscillations (Mie plasmons) in simple metal clusters~\cite{HeePRL87}. In a simple picture, the plasmon can be considered as a coherent oscillation of the valence electron cloud against the charged ionic background in a spherical metallic drop. Within the limit $r_{\rm cluster} \ll \lambda$, only the dipole mode couples to the laser field and may be excited resonantly with high cross section~\cite{Yab96,TigCPL96,SchEPJD99b}. A simple estimate for the spectral position of the Mie plasmon~\cite{Mie08} for a spherical metallic drop is $\omega_{\rm Mie}=\sqrt{e \,\rho_i /3 \varepsilon_0 \,m_e}$, which solely depends on the charge density of the ionic background $\rho_i$, while $\varepsilon_0$, $e$, and $m_e$ denote the vacuum permittivity, the elementary charge and the electron mass. In reality, additional effects like electronic spill out, polarization of core electrons, electronic quantum confinement, and cluster deformation lead to shifts and spectral fragmentation of the collective resonance~\cite{BraRMP00,Hee93,Cal00}. For our purposes, however, the plasmon can be considered as a single resonance with an energy $\hbar\omega_{\rm res}$ of a few eV, typically well above the photon energy of near-infrared (NIR) laser fields, and a lifetime $\tau_{\rm res}$ of a few fs.

In the weak field regime, the spectral position, width, and oscillator strength of the plasmon in gas phase clusters may be determined by spectral measurements of photofragmentation or photoionization cross sections with tunable narrow-band lasers~\cite{HeePRL87,TigCPL96,TigPRA93,SchEPJD99b}. For deposited cluster also two-photon photoemission has been applied~\cite{SchAPB01}. When aiming at the strong-field domain, a time-domain analysis of plasmon oscillations with sub-cycle resolution becomes desirable, as ionization and cluster explosion result in transient optical properties. This has not been achieved yet, though the time-dependence of the plasmon mode itself is routinely exploited in strong-field experiments. For example, the expansion-induced red-shift of the plasmon frequency occurring on the time scale of nuclear motion makes it possible to establish a plasmon resonance with NIR laser pulses, e.g. in a pump-probe setup~\cite{FenRMP10,SaaJPB06,DoePRL05,DoePRA06}. A sub-fs analysis of plasmon oscillations may thus be a challenging stepping stone towards a deeper understanding of collective processes in intense laser fields.

In the present study we propose a possible route towards such ultrafast time-resolved analysis. To this end we develop and theoretically explore a scheme to excite and image plasmon oscillations in expanding simple metal clusters with sub-fs resolution by nonlinear spectral interferometry with few-cycle laser pulses. The excitation and analysis of the plasmons is utilized by a pair of few-cycle pulses, where the plasmon oscillations are induced by the first pulse and probed with the second one. This enables a time-resolved mapping of the coherent electron motion by scanning the delay between the few-cycle pulses with interferometric accuracy. The performance and accuracy of this approach is demonstrated with semiclassical Vlasov simulations on the model system Na$_{147}$, where we show the possibility to resolve the dependence of the plasmon lifetime on the intensity of the pump pulse, which is a truly nonlinear effect. A key finding of our analysis is that experimentally accessible observables like ionization or pulse depletion can be used for the analysis, even if the system response is nonlinear.

The manuscript is structured as follows. Sec.~\ref{sec:specint_theory} introduces the strategy and approximations of our spectral interferometry approach and presents benchmark results for a simplified oscillator model. Sec.~\ref{sec:vlasovresults} discusses the semiclassical Vlasov approach and presents the numerical results on the time-resolved imaging of plasmon oscillations for representative cases. Conclusions and an outlook are given in Sec.~\ref{sec_conclusions}.

\section{Spectral interferometry}
\label{sec:specint_theory}
Spectral interferometry is the basis of Fourier transform spectroscopy and has also been recognized as a powerful tool for coherent control of molecular reactions with light~\cite{BruCPL86}. It has been successfully applied to various scenarios, such as the control of electronic wave packet motion in atoms~\cite{BouEPJD98,BlaPRL97}, rovibrational nuclear dynamics~\cite{SchJCP91}, and quantum interferences in molecules~\cite{KatPRL09}.

For a brief introduction of key aspects relevant for the application of spectral interferometry to collective processes and the derivation of important approximations, a simplified case in linear response is considered. A suitable benchmark scenario is the excitation of a damped harmonic oscillator by a pair of linearly polarized few-cycle laser pulses. Based on that, we discuss and test two approximations for the total energy absorption as function of pulse delay, which are key to the retrieval of the to-be-analyzed dynamical properties.
\subsection{Basic equations}
We begin with the electric field of a single pulse
\begin{equation}
E_{0}(t)=\frac{1}{2}E_{\rm env}(t)e^{-i(\omega_0t+\varphi_{\rm CE})}+c.c.,
\end{equation}
where $\omega_{0}$ is the angular frequency of the carrier wave, $\varphi_{\rm CE}$ is the carrier-envelope phase (CEP) and $E_{\rm env}(t)$ is the real-valued temporal field envelope. The total field of the pulse pair $E(t)=E_{0}(t)+E_{0}(t-\Delta t)$ then reads
\begin{equation}
E(t)=\frac{e^{-i\varphi_{\rm CE}}}{2}\left[E_{\rm env}(t) e^{-i\omega_0t}+E_{\rm env}(t-\Delta t)e^{-i\omega_0(t-\Delta t)}\right]+c.c.,
\end{equation}
where $\Delta t$ is the pulse delay. Note that we require pulses with the same CEP, which may be realized by a Michelson or Mach-Zehnder interferometer setup.
The total field in the Fourier domain
$E(\omega)=\int_{-\infty}^{\infty} E(t)e^{i\omega t}dt$ has the form
\begin{equation}
E(\omega)=\frac{1+e^{i\omega\Delta t}}{2}\left[e^{-i\varphi_{\rm CE}} E_{\rm env}(\omega-\omega_0)+ e^{i\varphi_{\rm CE}} E_{\rm env}(\omega+\omega_0)\right],
\end{equation}
where $E_{\rm env}(\omega)$ is the Fourier representation of the envelope. The net energy absorption of the oscillator driven by the light field $E(t)$ can be expressed in the Fourier domain 
 as an integral over the product of the spectral intensity $I(\omega)=1/2\pi\, |E(\omega)|^2$ and the absorption cross section of the oscillator $\sigma(\omega)$ via
\begin{equation}
W_{\rm abs}(\Delta t)=\frac{1}{2\pi}\int_{-\infty}^{\infty} E(\omega)E^*(\omega)\sigma(\omega) d\omega. 
\end{equation}
 Assuming $E_{\rm env}(\omega-\omega_0)E^*_{\rm env}(\omega+\omega_0)=0$, i.e. a sufficiently well peaked spectrum of the envelope function, and
exploiting that $\sigma(\omega)=\sigma(-\omega)$ for a classical oscillator we can write
\begin{equation}
W_{\rm abs}(\Delta t)=\frac{1}{2\pi}\int_{-\infty}^{\infty} \left( 1+\cos\omega \Delta t\right) |E_{\rm env}(\omega-\omega_0)|^2\sigma(\omega) d\omega.
\end{equation}
By this reformulation all spectral contributions around \mbox{$\omega=-\omega_0$} have been shifted to positive frequencies. It should further be noted that the CEP has dropped out. For sufficiently weak damping of the oscillator we can now express the cross section by a normalized Lorentzian and obtain
\begin{eqnarray}
W_{\rm abs}(\Delta t)=\frac{\gamma}{\pi}\int_{-\infty}^{\infty} \frac{\left( 1+\cos\omega \Delta t\right) |E_{\rm env}(\omega-\omega_0)|^2/2\pi}{(\omega_{\rm res}-\omega)^2+\gamma^2} d\omega,
\label{eq:wabs_full}
\end{eqnarray}
where $\omega_{\rm res}$ and $\gamma$ specify the frequency and spectral width (inverse lifetime) of the oscillator resonance. The numerator in the integral is the intensity spectrum of the double-pulse laser field $I(\omega)$. In this formulation of the spectral intensity all significant spectral contributions have been shifted to the range of positive frequencies. The usual relation $\int_{-\infty}^{\infty} I(\omega)d\omega =\int_{-\infty}^{\infty} E^2(t)dt$ is still valid. The double-pulse spectrum $I(\omega)$
contains the intensity spectrum of a single pulse
\begin{equation}
I_0(\omega)=|E_{\rm env}(\omega-\omega_0)|^2/4\pi
\label{eq:Isingle}
\end{equation}
times a spectral modulation with $2(1+\cos \omega\Delta t)$. This tunable interference of the pulses in the Fourier domain (cf. insets of Fig.~\ref{fig:model_spectra}a) is the key feature exploited in spectral interferometry. For extracting the deeper physical meaning, the integral in Eq.\,(\ref{eq:wabs_full}) will now be considered in detail for two approximations.

\subsection{Linear spectrum approximation (LSA)}
Assuming an oscillator resonance well inside the laser spectrum (detuning up to the spectral width of the laser pulse), $I_{0}(\omega)$ can be linearized around $\omega_{\rm res}$ by
\begin{equation}
I_{0}^{\rm LSA}(\omega)=\alpha+(\omega-\omega_{\rm res})\beta,
\end{equation}
with $\alpha=I_{0}(\omega_{\rm res})$ and  $\beta={\partial I_{0}(\omega)}/ {\partial  \omega}|_{\omega=\omega_{\rm res}}$. Carrying out the integral in Eq.\,(\ref{eq:wabs_full}) within this linear spectrum approximation (LSA) yields an absorption of
\begin{equation}
W_{\rm abs}^{\rm LSA}(\Delta t)=2\alpha\left(1+\sqrt{1+c_{\rm LSA}^2} \,\cos \left (\omega_{\rm res}\Delta t+\varphi_{\rm LSA} \right )e^{-\gamma \Delta t}\right),
\label{eq:wabs_lin}
\end{equation}
where $c_{\rm LSA}=\tan \varphi_{\rm LSA}=\beta\gamma/\alpha$.

In the special case of an oscillator resonance at the maximum of $I_0(\omega)$, i.e. for $\beta=0$, this reduces further to
\begin{equation}
W_{\rm abs}^{\rm LSA}(\Delta t)=2\alpha\left( 1+ \cos \left (\omega_{\rm res}\Delta t \right )e^{-\gamma \Delta t}\right ).
\label{eq:wabs_lin_max}
\end{equation}
Hence, the delay-dependent absorption exhibits a static offset plus a damped periodic oscillation with the eigenfrequency and lifetime of the oscillator. Most importantly, for vanishing CEP the oscillations even provide a direct time-domain image of the dipole velocity of the oscillator after the first pulse, which turns out to be the key to our imaging scheme. Apart from a phase shift by $\varphi_{\rm LSA}$ and a scaled modulation amplitude (both can be compensated for a known laser spectrum), this picture even persists if the resonance is located in the wings of the laser spectrum ($\beta \neq 0$), cf. Eq.\,(\ref{eq:wabs_lin}). It will be shown later that these modifications are small even for considerable detuning of the order of the spectral width of the laser pulse.

\subsection{Constant cross-section approximation (CCA)}
\label{sec:specint_theory:cca}
As the second limiting case we consider a fully off-resonant excitation, i.e. $\omega_{\rm res}$ far outside the laser spectrum. The cross section can then be approximated with the value at the carrier frequency $\sigma_0=\sigma(\omega_0)$. The absorption integral within this constant cross section approximation (CCA) becomes
\begin{eqnarray}
W_{\rm abs}^{\rm CCA}(\Delta t)=2\sigma_0\int_{-\infty}^{\infty} (1+\cos \omega\Delta t)I_0(\omega)d\omega.
\label{eq:wabs_stat}
\end{eqnarray}
Applying the convolution theorem one finds
\begin{eqnarray}
W_{\rm abs}^{\rm CCA}(\Delta t)&=&\sigma_0\int_{-\infty}^{\infty}E^2_{\rm env}(t)dt\\
&+&\sigma_0\cos \omega_0 \Delta t\int_{-\infty}^{\infty}E_{\rm env}(t)E_{\rm env}(t-\Delta t)dt \nonumber\\
&=&\sigma_0 \underbrace{\left[ 2 F_0+ A_{\rm env}(\Delta t) \cos \omega_0 \Delta t \right ]}_{F(\Delta t)}
\label{eq:wabs_stat2}
\end{eqnarray}
which contains the single pulse fluence $F_0=\frac{1}{2}\int_{-\infty}^{\infty}E^2_{\rm env}(t)dt$ and the autocorrelation of the envelope function  $A_{\rm env}(\Delta t)=\int_{-\infty}^{\infty} E_{\rm env}(t)E_{\rm env}(t-\Delta t)dt$. The term in square brackets in Eq.\,(\ref{eq:wabs_stat2}) can be identified with the fluence $F(\Delta t)$ of the double pulse laser field. Hence, a non-resonant excitation results in an absorption signal proportional to the field fluence, which can be exploited for identification of such scenario.

\subsection{Simple examples with gaussian few-cycle pulses}
We now demonstrate the performance of the above approximations for representative benchmark examples with a damped oscillator. Considering gaussian pulses, the envelope function can be written as $E_{\rm env}(t)=\hat E e^{-t^2/\tau^2}$, where $\hat E$ is the single pulse field amplitude and $\tau$ is the pulse width. The latter can be related to the full width at half maximum of the intensity via $\tau=\tau_{\rm FWHM}/\sqrt{2\ln 2}$. With the envelope function in the Fourier domain $E_{\rm env}(\omega)=\hat E\tau\sqrt{\pi}\,e^{-\omega^2\tau^2/4}$ the spectral intensity of the single pulse is
\begin{eqnarray}
I_0(\omega)=\hat E^2\frac{\tau^2}{4} \, e^{-(\omega-\omega_0)^2\tau^2/2},
\end{eqnarray}
which yields $c_{\rm LSA}=\tan \varphi_{\rm LSA}=-\tau^2\gamma (\omega_{\rm res}-\omega_0)$ for the linear spectrum approximation in Eq.\,(\ref{eq:wabs_lin}).

\begin{figure}[h!]
\centering
  \includegraphics[width=8.2cm]{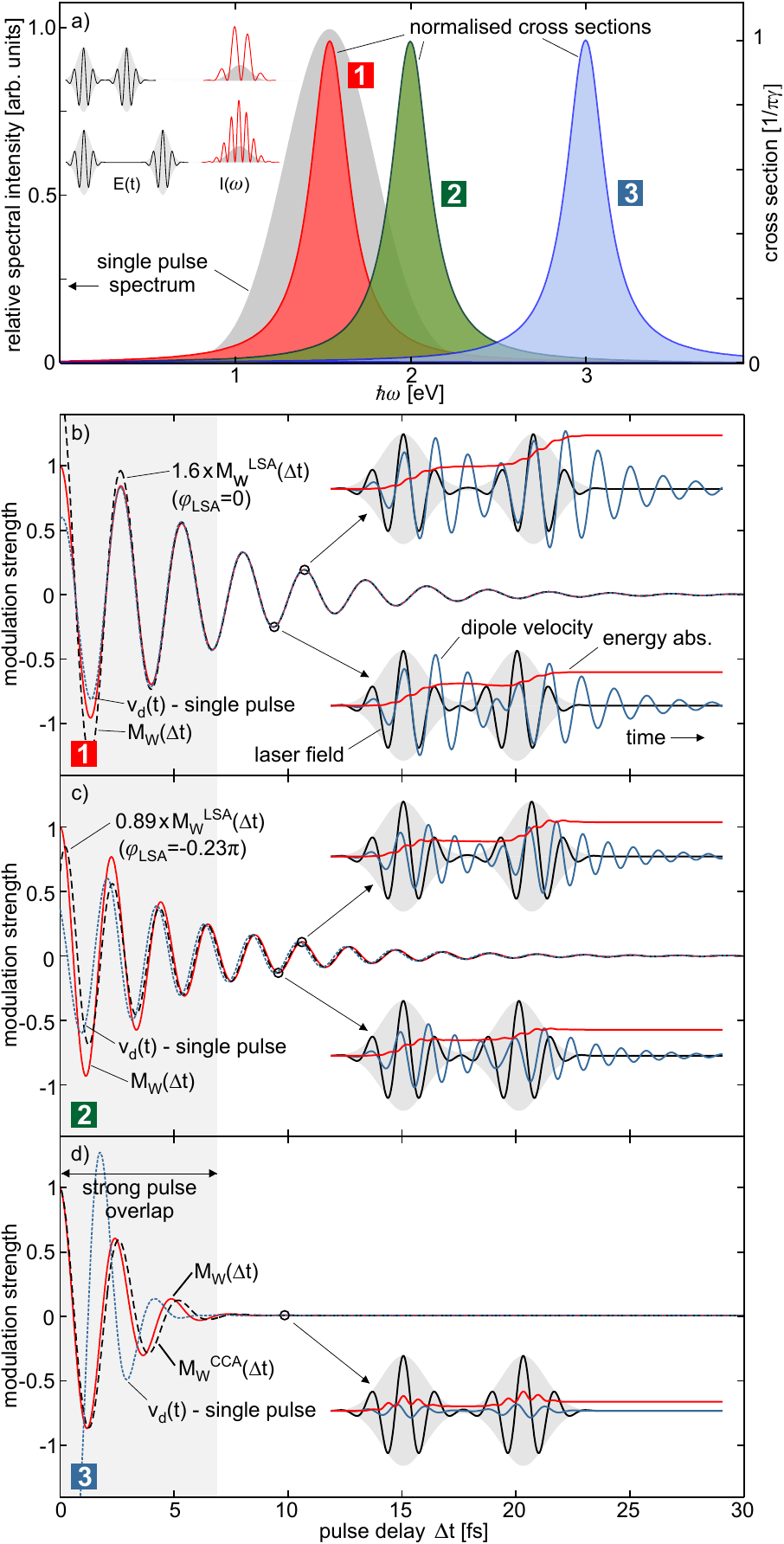}
  \caption{Spectral interferometry results with gaussian 3\,fs pulses ($\hbar\omega_0=1.54\,$eV) for three  Lorentz oscillator settings. (a) single pulse spectral intensity (gray) and normalised oscillator absorption cross sections for $\omega_{\rm res}=\omega_0$, $1.3\omega_0$, and $2\omega_0$ (red, green, blue) for $\gamma^{-1}=5\,$fs lifetime; insets show electric field evolutions (black) and resulting spectral intensity distributions (red) for different delays; (b-d) delay-dependent modulation signals $M_W(\Delta t)$ (red) compared to results in linear spectrum approximation (LSA) or constant cross section approximation (CCA); corresponding $\varphi_{\rm LSA}$ values as indicated; dashed blue curves show the pump-only dipole velocity signal $v_{\rm d}(t)$; insets depict evolutions of the laser field (black), dipole velocity (blue), and absorbed energy (red) for indicated delays.\label{fig:model_spectra}}
\end{figure}

In the next step, optical excitation with few-cycle pulses (\mbox{$\tau_{\rm FWHM}=3\,$fs}) at 800\,nm carrier-wavelength \mbox{($\hbar\omega_0$=1.54\,eV)} with zero CEP is considered. Exemplarily, we study three different oscillator configurations with resonances \mbox{$\omega_{\rm res}=\omega_0$}, 1.3$\omega_0$, and $2\omega_0$. For all cases a fixed damping constant of $\hbar\gamma=0.13\,$eV (this corresponds to a lifetime $\tau_{\rm res}=5\,$fs) is assumed. These examples reflect (i) fully resonant, (ii) near-resonance, and (iii) nonresonant excitation. Normalized cross sections are shown Fig.~\ref{fig:model_spectra}a in relation to the single pulse intensity spectrum.

Lets now consider some specific observable $A$, which may be absorption, mean electron energy, or electron yield in a real cluster experiment. The delay dependence of $A(\Delta t)$ can then be expressed by the dimensionless modulation parameter
\begin{equation}
M_A(\Delta t)=\frac{A(\Delta t)}{A(\Delta t' \gg \tau_{res})}-1,
\end{equation}
which vanishes for pulse delays much larger than the oscillator lifetime. The modulation parameter for the absorbed energy with the full expression in Eq.\,(\ref{eq:wabs_full}) then follows as $M_W(\Delta t)$. This function is used to benchmark the modulation parameters $M_W^{\rm LSA}(\Delta t)$ and $M_W^{\rm CCA}(\Delta t)$ obtained from the approximated energy absorptions
in Eq.\,(\ref{eq:wabs_lin}) or Eq.\,(\ref{eq:wabs_stat}), respectively. The resulting evolutions of $M_W(\Delta t)$ as function of pulse delay (solid red curves) are shown in Figs.~\ref{fig:model_spectra}b-\ref{fig:model_spectra}d in relation to the applicable approximation $M_W^{\rm LSA}(\Delta t)$ or $M_W^{\rm CCA}(\Delta t)$ (dashed, approximation as indicated).

For the fully resonant and the near-resonant scenario (Figs.~\ref{fig:model_spectra}b and \ref{fig:model_spectra}c), the predictions of the linear spectrum approximation $M_W^{\rm LSA}(\Delta t)$ (rescaled in amplitude as indicated) are in good agreement with the exact results $M_W(\Delta t)$ and show damped oscillations with high contrast. A closer comparison of the rescaled $M_W^{\rm LSA}(\Delta t)$ signals with $M_W(\Delta t)$ shows, that the modulations appear in phase and with the oscillator frequency and decay time for $\Delta t > 6\,$fs, i.e., outside the region with strong pulse overlap. Deviations for smaller delays reflect signal disturbance by higher order interference terms that are negligible outside the pulse overlap region.

For selected delays, time-domain evolutions of the laser field, dipole velocity, and absorbed energy are sketched as insets and provide a direct image of the dynamics that is mapped into the modulation parameters.
Focussing on the upper time-domain plot in Fig.~\ref{fig:model_spectra}b, a resonant dipole oscillation is excited by the pump pulse. The fully resonant nature is reflected by the fact, that the dipole velocity and the laser field are in phase. For the chosen delay, the probe pulse arrives in phase with the dipole velocity (constructive interference) and the oscillator motion can be strongly amplified. The energy gain from the probe clearly exceeds the absorption from the pump pulse due to advantageous coherent superposition of remaining coherent pump-induced motion and the probe excitation.

In contrast to that, the lower cartoon in Fig.~\ref{fig:model_spectra}b shows an example for destructive interference, where the probe pulse even stops and reverses the oscillator motion and the net energy gain from the probe is substantially reduced. The oscillation of the modulation parameter with pulse delay thus reflects the alternation of amplification and suppression of energy absorption from the probe pulse due to the residual oscillator motion. In turn, the decrease of the modulation amplitude with delay indicates the decay of the pump-pulse-induced oscillator motion. For long delays ($\Delta t\gg \tau_{\rm res}$) the interaction with the probe pulse is incoherent, as no phase-sensitive excitation remains at the time of its arrival.

In the fully resonant scenario ($\beta=0$) in Fig.~\ref{fig:model_spectra}b, the modulation signals directly image the pump-induced motion of the oscillator. This can be inferred from the close-to-perfect agreement of $M_W(\Delta t)$ and the rescaled $M_W^{\rm LSA}(\Delta t)$ with the dipole velocity $v_{\rm d}(t)$ for pump-only excitation (blue dashed curve in Fig.~\ref{fig:model_spectra}b). Outside the pulse overlap region, signals are in phase and show the same relative amplitude evolution. This direct mapping of oscillator motion into the modulation parameter is the heart of our imaging scheme.

A similar set of time-domain examples is shown for the near-resonance scenario in Fig.~\ref{fig:model_spectra}c (see insets). The general trends and the modulation effects are very similar to the fully resonant case. It should be emphasized that the modulation signals (Fig.~\ref{fig:model_spectra}c) are now phase shifted to the pump-induced dipole velocity by a small phase offset of the order of $\varphi_{\rm LSA}$. The latter could be determined by fitting the LSA formula from Eq.~(\ref{eq:wabs_lin}) to a given $M_W(\Delta t)$ signal outside the pulse overlap region or directly from the laser spectrum. However, even for detunings as large as the width of the pulse spectrum as in our example, the phase effect is small and may be neglected to first order. Under this approximation, the modulation signal can be interpreted as a direct image of the dipole velocity after the pump pulse, i.e. the time-evolution of the oscillator motion can be retrieved. Further note that our results have been obtained for zero CEP of the pulses. While the modulation parameter $M_{W}(\Delta t)$ is CEP-independent, the CEP will occur as a direct phase offset in the dipole velocity. Hence, in the general case the dipole velocity can be reconstructed by $v_{\rm d}(t)\propto \cos(\omega_{\rm res}t+\varphi_{\rm CE})e^{-\gamma t}$ up to constant factor, where $\omega_{\rm res}$ and $\gamma$ are obtained from the LSA fit, and time is measured with respect to the pulse peak.

Coming back to the nonresonant scenario in Fig.~\ref{fig:model_spectra}d, there remains no notable excitation after the pump pulse. The modulation parameter $M_W(\Delta t)$ almost exclusively images the trivial interference-induced fluence variations (interferometric autocorrelation) of the two pulses at small delays, as can be inferred from comparison to the $M_W^{\rm CCA}(\Delta t)$ data, cf. Eq.~(\ref{eq:wabs_stat2}). This behavior allows one to identify a nonresonant scenario.

\section{Imaging plasmons in simple-metal clusters}
\label{sec:vlasovresults}
We now turn to the application of the spectral interferometry scheme to the nonlinear response domain and present a
theoretical case study on simple-metal clusters. To describe the strong-field induced excitation and ionization dynamics of the model system Na$_{147}$ we apply a semiclassical time-dependent density-functional approach on the Vlasov level that has been used for strong-field laser-cluster interactions previously~\cite{FenRMP10,FenEPJD04,FenPRL07a}. As test systems we consider the cluster ground state and an expanded configuration to study the effect of the spectral position of the plasmon resonance. For the expanded system we further analyze the dependence of the cluster response on pulse intensity. It is demonstrated that the spectral interferometry analysis is capable of extracting dynamical observables with sub-fs time resolution and can be performed with experimentally accessible observables.

\subsection{Semiclassical Vlasov approach}
The applied Thomas-Fermi-Vlasov approach is a semiclassical approximation to time-dependent density-functional theory and describes the electron dynamics in terms of a continuous one-particle electron phase-space density $f({\bf r},{\bf p},t)$. The time evolution of $f({\bf r},{\bf p},t)$ is derived from the quantal mean-field dynamics by applying the well-known $\hbar\rightarrow 0$ expansion~\cite{BerPR88,DomAP97a,PlaPRA00,FenEPJD04,FenLNP08_}. In lowest order this yields the Vlasov equation
\begin{eqnarray}
\frac{\partial f({\bf r},{\bf p},t)}{\partial t}=\nabla_{\bf p}f({\bf r},{\bf p},t) \cdot \nabla_{\bf r}V_{\rm eff}({\bf r},t)- \frac{{\bf p}}{m} \cdot \nabla_{\bf r}f({\bf r},{\bf p},t),
\label{eq:3B3_vlasov}
\end{eqnarray}
which is the equation of motion for the electronic degrees of freedom. The effective mean-field potential $V_{\rm eff}({\bf r},t)$ follows from variation of the density-dependent total potential energy functional with respect to the density as
\begin{equation}
V_{\rm eff}({\bf r})=\sum_{\rm i} V_{\rm ion}({\bf r}-{\bf  R}_i(t))+V_{\rm Har}({\bf r})+V_{\rm xc}({\bf r})+e\,{{\bf  E}}(t) \cdot{\bf r},
\label{eq:Veff}
\end{equation}
containing the sum over the ion potentials for the present configuration ${\bf R}_i(t)$, the electron
Hartree potential $V_{\rm Har}$, the exchange-correlation potential $V_{\rm xc}$ in local-density approximation (LDA) [we use the form given in Ref.~\cite{GunPRB76}], and the interaction with the laser field ${\bf E}(t)$ in dipole approximation. The Hartree term and the exchange-correlation potential are calculated self-consistently from the actual total electron density \mbox{$n_e({\bf r},t)=\int d^3{\bf p} f({\bf r},{\bf p},t)$}. To avoid the expensive propagation of strongly localized states, only valence electrons are treated explicitly. The interaction with nuclei and core electrons is described by a local pseudopotential $V_{\rm ion}(r)$ for sodium~\cite{FenEPJD04}. Classical dynamics is assumed for the ionic motion.

Though Eq.\,(\ref{eq:3B3_vlasov}) is of classical nature, limited quantum effects such as exchange and correlation in LDA are kept in the effective potential and the initial conditions for the distribution function. The latter is determined from the Thomas-Fermi ground state according to $ \displaystyle f^0({\bf r},{\bf p})=2/(2\pi\hbar)^3\Theta(p_{_{\rm F}}({\bf r})-p), $ where $\Theta(x)$ is the Heaviside function, $p_{_{\rm F}}({\bf r})=\sqrt{2m[\mu-V_{\rm eff}({\bf r})]}$ is the local Fermi momentum, and $\mu$ is the chemical potential to fix the electron number. Propagation of the initial distribution $f^0({\bf r},{\bf p})$ by Eq.~(\ref{eq:3B3_vlasov}) constitutes the Thomas-Fermi-Vlasov dynamics. For solving the ground-state problem and the time-propagation we apply the test particle method in a parallel particle-mesh code with an iterative multigrid Poisson solver. Further technical details are described elsewhere~\cite{FenEPJD04,KoePRA08,FenLNP08_}.

\subsection{Configurations of the model systems}
\begin{figure}[h!]
\centering
  \includegraphics[width=8cm]{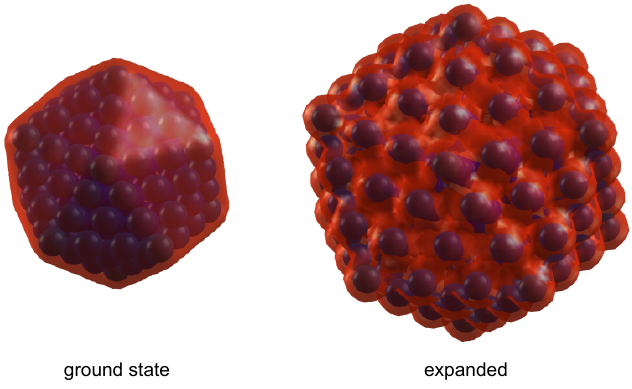}
  \caption{Calculated ion and electron distributions of icosahedral Na$_{147}$ in the ground state and in the expanded configuration (as indicated), both in the electronic ground state (see text). The 3D images show ion positions as blue spheres and electron isodensity surfaces for $n´_e=4\times10^{21}$cm$^{-3}$. The expanded configuration follows from scaling the ground state ion geometry by a factor of 1.55, resulting in a classical Mie plasmon energy of about 1.5\,eV. The ion root-mean-square radii are 8.8 and 13.6$\,{\rm \AA}$, respectively.}
  \label{fig:configurations}
\end{figure}

We consider two different configurations of our test system Na$_{147}$. The first one is the ground state with fully relaxed ion configuration. This yields an icosahedral structure with highly delocalized valence electrons with a mean density close to the bulk electron density of sodium (cf. Fig.~\ref{fig:configurations}). The ground state has a well localized plasmon resonance at roughly 3\,eV and will be used as an off-resonant target.

As a resonant configuration we consider an artificially expanded cluster, generated by rescaling of the ionic ground state geometry. The scaling factor is chosen such that the background ion density yields a classical Mie plasmon in resonance with a laser field at 800\,nm. Also for this ion configuration, the electronic ground state is calculated (cf. Fig.~\ref{fig:configurations}). Note that such expansion can be realized experimentally by applying an earlier activation pulse. Though this more realistic scheme would result in an expanded system that is charged and thermally excited, the electronic ground state is used here for convenience and clarity.

\subsection{Nonlinear spectral interferometry results}
\begin{figure*}[t!]
\centering
  \includegraphics[width=18cm]{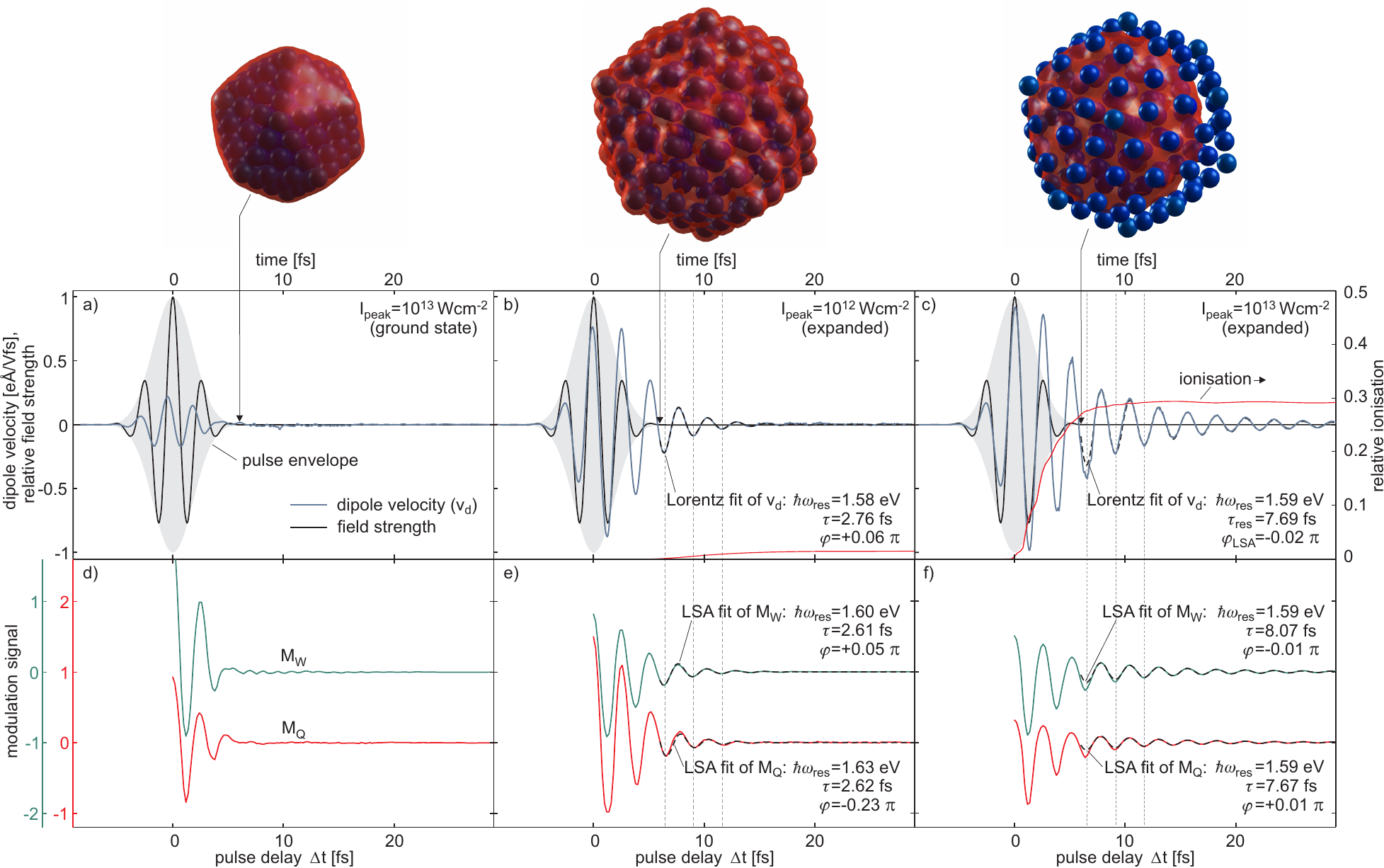}
  \caption{Plasmon dynamics and spectral interferometry analysis of Na$_{147}$ under intense 3\,fs few-cycle pulses at $\hbar\omega_0=1.54\,$eV as calculated within the semiclassical Vlasov approach (cluster configurations and laser peak intensities as indicated). (a-c) time-evolutions of the laser field, dipole velocity (relative to the peak field amplitude) and cluster ionization for pump-only excitation; Lorentz fits of the dipole velocity for $t>6\,$fs; 3D snapshot images at $t=6$\,fs. (d-f) systematic delay scans of the modulation signals of the absorption and cluster ionization [$M_W(\Delta t)$, $M_{Q}(\Delta t)$] and corresponding LSA fits for pump-probe excitation.}
  \label{fig:plasmon_scan}
\end{figure*}

We now consider laser excitation of the above described cluster configurations by a pair of intense 3\,fs few-cycle pulses with zero carrier-envelope phase at 800\,nm. As we are aiming at the analysis of the pump-pulse-induced collective electron motion we begin with an inspection of key observables for excitation with the pump pulse only, see upper panels of  Fig.~\ref{fig:plasmon_scan}.

As a nonresonant reference case, pump-only excitation of the unexpanded cluster is considered at laser intensity \mbox{$I_{\rm peak}=10^{13}$\,W/cm$^2$}, for selected time-dependent observables see Fig.~\ref{fig:plasmon_scan}a. The fully off-resonant excitation yields only very low energy absorption and the cluster dipole response is a nearly  instantaneous polarization that directly follows the laser field. The dipole velocity (blue curve) is phase shifted by $\pi/2$ with respect to the field oscillation of the laser pulse (black curve) and nearly vanishes immediately after the pulse. The ion and electron distribution of the cluster after the laser pulse are basically equal to the initial state (cf. Fig.~\ref{fig:plasmon_scan}a).

In addition, two resonant scenarios are depicted in Figs.~\ref{fig:plasmon_scan}b and \ref{fig:plasmon_scan}c, showing the response of the expanded cluster configurations for pump-only excitation with intensities $I_{\rm peak}=10^{12}$ and $10^{13}$ W/cm$^2$, respectively. In both cases, strong plasmon oscillations are excited, as is reflected in the oscillations of the dipole velocity after the end of the laser pulse. Note that the eigenfrequencies determined from damped oscillator fits are very similar ($\hbar\omega_{\rm res}\approx 1.6\,$eV) for both intensities.

Most interestingly, for the more intense pump pulse, the plasmon oscillations exhibits a substantially longer lifetime of $\tau_{\rm res}=7.7\,$fs over the result for the lower intensity scenario of 2.8\,fs, which is a nonlinear effect of the excitation process. This behavior can be understood by considering the higher cluster ionization achieved with the stronger pulse, see the relative ionization (red curves) in Figs.~\ref{fig:plasmon_scan}b and \ref{fig:plasmon_scan}c. For the higher intensity, the remaining electron cloud is confined to a volume smaller than the ionic frame of the cluster (compare 3D plots in Figs.~\ref{fig:plasmon_scan}b and \ref{fig:plasmon_scan}c). When oscillating with not-too-large amplitude, the electron cloud traverses only the inner cluster region where the ion background potential is nearly harmonic, which effectively reduces plasmon dephasing by interactions with the anharmonic background potential near the cluster surface. Hence, the nonlinear effect yields an intensity dependent increase of the plasmon lifetime, while the plasmon oscillation itself remains an effectively linear phenomenon. This is analogous to the nonlinear index of refraction (effective linear property) of gases or solids.

Having understood the main physics of the three test cases, we now apply the spectral interferometry analysis. Therefore the cluster response for pump-probe excitation is calculated for systematic delay-scans. For each run the total energy absorption and the final ionization is recorded for 200 different delays. Modulation signals $M_{W}(\Delta t)$ and  $M_{Q}(\Delta t)$ are then calculated from the absorption and the final ionization as function of pulse delay, see Figs.~\ref{fig:plasmon_scan}d-\ref{fig:plasmon_scan}f. Before inspecting their evolution in detail, we have to motivate the use of these parameters for our spectral interferometry analysis, which was formulated for a linear absorption cross section. The key idea is that the modification of the probe-pulse induced absorption and ionization is changed only weakly by the residual coherent motion of the plasmon, which allows one to linearize the absorption and ionization as function of the residual dipole velocity amplitude. This approximation will be shown to be applicable outside the pulse overlap region.

Focussing on the nonresonant case (Fig.~\ref{fig:plasmon_scan}d) first, both the $M_W(\Delta t)$ and $M_Q(\Delta t)$ signals show oscillations only for very short delays within the region of strong pulse overlap. This is the typical signature of a nonresonant excitation, in full agreement with the behavior predicted within the constant cross section approximation discussed in Sec.~\ref{sec:specint_theory:cca}. The signals thus show an image of the interference of the pulses, though signals may not be exactly proportional to the field fluence because of the nonlinear system response. The nonresonant nature, however, can be unambiguously be identified from the modulation signals.

For the resonant scenarios, the modulation parameters show strong oscillations well beyond the pulse autocorrelation, see Figs.~\ref{fig:plasmon_scan}e and \ref{fig:plasmon_scan}f. Focussing on the region outside the pulse overlap, a nearly perfect mapping of the pump-pulse induced dipole velocity into the modulation signals is observed, i.e. $M_W(\Delta t)\approx M_Q(\Delta t) \propto v_{\rm d}(t)$, where $v_{\rm d}(t)$ is the corresponding dipole velocity for pump-only excitation. In particular, the different lifetimes are clearly resolved and the damped oscillator fits of the dipole signals and the LSA fits of $M_W(\Delta t),M_Q(\Delta t)$ yield frequencies and lifetimes that agree within a few percent. The extracted phase offsets are close to zero, as expected for fully resonant excitation.
In should be noted that the modulation signals can be well described by the fit functions, except for short delays in the high intensity scenario, where deviations due to nonlinearities can be found in the delay range $\Delta t=7\ldots 9\,$fs, see Fig.~\ref{fig:plasmon_scan}f. Otherwise the assumption of a linearized response seems to be well justified.

These results support the claim that nonlinear spectral interferometry with few-cycle pulses opens up an avenue towards imaging of strong-field induced plasmons in metal-clusters with sub-fs time resolution. In addition, as the to-be-scanned delay window can be as narrow as a few tens of fs, our approach is expected to provide valuable insights into the time-evolution of plasmonic properties in expanding clusters, which would be of high interest for the whole area of nanoplasma science.

\section{Conclusions and outlook}
\label{sec_conclusions}
In conclusion, we have explored the applicability of spectral interferometry with few-cycle laser pulses as an imaging tool for ultrafast collective electron dynamics in the nonlinear response domain. Two simple approximations have been derived that can be used for fitting the modulation signals of energy absorption, ionization, or other (linearizable) observables. The performance of the method has been tested with a ``numerical experiment'' on simple-metal clusters, where the time-domain analysis of plasmon oscillations with sub-fs resolution at high modulation contrast has been demonstrated. Our results support that both the time-resolved plasmon motion as well as key properties such as frequency and lifetime can be accurately determined. Moreover, we have shown that strong-field effects, like the increase of the plasmon lifetime due to ionization-induced contraction of the electron cloud could be resolved. It should be emphasized that the accurate analysis of the dynamics has demonstrated by evaluating experimentally accessible observables, such as total ionization. The unprecedented time-resolution of our method promises new fundamental insights into the
strong-field induced dynamics in finite many particle systems.

\section{Acknowledgements}
Financial support by the Deutsche Forschungsgemeinschaft
within the SFB 652/2 and computer time provided by the HLRN computing center are gratefully acknowledged.

\footnotesize{
\bibliographystyle{rsc} 

\begin{mcitethebibliography}{49}
\providecommand*{\natexlab}[1]{#1}
\providecommand*{\mciteSetBstSublistMode}[1]{}
\providecommand*{\mciteSetBstMaxWidthForm}[2]{}
\providecommand*{\mciteBstWouldAddEndPuncttrue}
  {\def\EndOfBibitem{\unskip.}}
\providecommand*{\mciteBstWouldAddEndPunctfalse}
  {\let\EndOfBibitem\relax}
\providecommand*{\mciteSetBstMidEndSepPunct}[3]{}
\providecommand*{\mciteSetBstSublistLabelBeginEnd}[3]{}
\providecommand*{\EndOfBibitem}{}
\mciteSetBstSublistMode{f}
\mciteSetBstMaxWidthForm{subitem}
{(\emph{\alph{mcitesubitemcount}})}
\mciteSetBstSublistLabelBeginEnd{\mcitemaxwidthsubitemform\space}
{\relax}{\relax}

\bibitem[Rabitz \emph{et~al.}(2000)Rabitz, de~Vivie-Riedle, Motzkus, and
  Kompa]{RabSci00}
H.~Rabitz, R.~de~Vivie-Riedle, M.~Motzkus and K.~Kompa, \emph{Science}, 2000,
  \textbf{288}, 824\relax
\mciteBstWouldAddEndPuncttrue
\mciteSetBstMidEndSepPunct{\mcitedefaultmidpunct}
{\mcitedefaultendpunct}{\mcitedefaultseppunct}\relax
\EndOfBibitem
\bibitem[Rice and Zhao(2000)]{Ric00}
S.~A. Rice and M.~Zhao, \emph{Optical Control of Molecular Dynamics},
  Wiley-Interscience, New York, 2000\relax
\mciteBstWouldAddEndPuncttrue
\mciteSetBstMidEndSepPunct{\mcitedefaultmidpunct}
{\mcitedefaultendpunct}{\mcitedefaultseppunct}\relax
\EndOfBibitem
\bibitem[Nuernberger \emph{et~al.}(2007)Nuernberger, Vogt, Brixner, and
  Gerber]{NuePCCP07}
P.~Nuernberger, G.~Vogt, T.~Brixner and G.~Gerber, \emph{Phys. Chem. Chem.
  Phys.}, 2007, \textbf{9}, 2470\relax
\mciteBstWouldAddEndPuncttrue
\mciteSetBstMidEndSepPunct{\mcitedefaultmidpunct}
{\mcitedefaultendpunct}{\mcitedefaultseppunct}\relax
\EndOfBibitem
\bibitem[Shapiro and Brumer(2003)]{Sha03}
M.~Shapiro and P.~Brumer, \emph{Principles of the Quantum Control of
  Mole\-cular Processes}, Wiley-Interscience, Hoboken, 2003\relax
\mciteBstWouldAddEndPuncttrue
\mciteSetBstMidEndSepPunct{\mcitedefaultmidpunct}
{\mcitedefaultendpunct}{\mcitedefaultseppunct}\relax
\EndOfBibitem
\bibitem[Judson and Rabitz(1992)]{JudPRL92}
R.~S. Judson and H.~Rabitz, \emph{Phys. Rev. Lett.}, 1992, \textbf{68},
  1500\relax
\mciteBstWouldAddEndPuncttrue
\mciteSetBstMidEndSepPunct{\mcitedefaultmidpunct}
{\mcitedefaultendpunct}{\mcitedefaultseppunct}\relax
\EndOfBibitem
\bibitem[Daniel \emph{et~al.}(2003)Daniel, Full, Gonz\'alez, Lupulescu, Manz,
  Merli, Vajda, and W\"oste]{DanSci03}
C.~Daniel, J.~Full, L.~Gonz\'alez, C.~Lupulescu, J.~Manz, A.~Merli, S.~Vajda
  and L.~W\"oste, \emph{Science}, 2003, \textbf{299}, 536\relax
\mciteBstWouldAddEndPuncttrue
\mciteSetBstMidEndSepPunct{\mcitedefaultmidpunct}
{\mcitedefaultendpunct}{\mcitedefaultseppunct}\relax
\EndOfBibitem
\bibitem[Krausz and Ivanov(2009)]{KraRMP09}
F.~Krausz and M.~Ivanov, \emph{Rev. Mod. Phys.}, 2009, \textbf{81}, 163\relax
\mciteBstWouldAddEndPuncttrue
\mciteSetBstMidEndSepPunct{\mcitedefaultmidpunct}
{\mcitedefaultendpunct}{\mcitedefaultseppunct}\relax
\EndOfBibitem
\bibitem[Brixner \emph{et~al.}(2004)Brixner, Krampert, Pfeifer, Selle, Gerber,
  Wollenhaupt, Graefe, Horn, Liese, and Baumert]{BriPRL04}
T.~Brixner, G.~Krampert, T.~Pfeifer, R.~Selle, G.~Gerber, M.~Wollenhaupt,
  O.~Graefe, C.~Horn, D.~Liese and T.~Baumert, \emph{Phys. Rev. Lett}, 2004,
  \textbf{92}, 208301\relax
\mciteBstWouldAddEndPuncttrue
\mciteSetBstMidEndSepPunct{\mcitedefaultmidpunct}
{\mcitedefaultendpunct}{\mcitedefaultseppunct}\relax
\EndOfBibitem
\bibitem[Weiner(2000)]{WeiRSI00}
A.~M. Weiner, \emph{Rev. Sci. Instr.}, 2000, \textbf{71}, 1929--1960\relax
\mciteBstWouldAddEndPuncttrue
\mciteSetBstMidEndSepPunct{\mcitedefaultmidpunct}
{\mcitedefaultendpunct}{\mcitedefaultseppunct}\relax
\EndOfBibitem
\bibitem[Keller(2003)]{KelN03}
U.~Keller, \emph{Nature}, 2003, \textbf{424}, 831--838\relax
\mciteBstWouldAddEndPuncttrue
\mciteSetBstMidEndSepPunct{\mcitedefaultmidpunct}
{\mcitedefaultendpunct}{\mcitedefaultseppunct}\relax
\EndOfBibitem
\bibitem[Winterfeldt \emph{et~al.}(2008)Winterfeldt, Spielmann, and
  Gerber]{WinRMP08}
C.~Winterfeldt, C.~Spielmann and G.~Gerber, \emph{Rev. Mod. Phys.}, 2008,
  \textbf{80}, 117\relax
\mciteBstWouldAddEndPuncttrue
\mciteSetBstMidEndSepPunct{\mcitedefaultmidpunct}
{\mcitedefaultendpunct}{\mcitedefaultseppunct}\relax
\EndOfBibitem
\bibitem[Pfeifer \emph{et~al.}(2005)Pfeifer, Walter, Winterfeldt, Spielmann,
  and Gerber]{PfeAPB05}
T.~Pfeifer, D.~Walter, C.~Winterfeldt, C.~Spielmann and G.~Gerber, \emph{Appl.
  Phys. B}, 2005, \textbf{80}, 277\relax
\mciteBstWouldAddEndPuncttrue
\mciteSetBstMidEndSepPunct{\mcitedefaultmidpunct}
{\mcitedefaultendpunct}{\mcitedefaultseppunct}\relax
\EndOfBibitem
\bibitem[Sansone \emph{et~al.}(2006)Sansone, Benedetti, Calegari, Vozzi,
  Avaldi, Flammini, Poletto, Villoresi, Altucci, Velotta, Stagira, Silvestri,
  and Nisoli]{SanSci06}
G.~Sansone, E.~Benedetti, F.~Calegari, C.~Vozzi, L.~Avaldi, R.~Flammini,
  L.~Poletto, P.~Villoresi, C.~Altucci, R.~Velotta, S.~Stagira, S.~D. Silvestri
  and M.~Nisoli, \emph{Science}, 2006, \textbf{314}, 443\relax
\mciteBstWouldAddEndPuncttrue
\mciteSetBstMidEndSepPunct{\mcitedefaultmidpunct}
{\mcitedefaultendpunct}{\mcitedefaultseppunct}\relax
\EndOfBibitem
\bibitem[Zeng \emph{et~al.}({2007})Zeng, Cheng, Song, Li, and Xu]{ZenPRL07}
Z.~Zeng, Y.~Cheng, X.~Song, R.~Li and Z.~Xu, \emph{{Phys. Rev. Lett.}}, {2007},
  \textbf{{98}}, 203901\relax
\mciteBstWouldAddEndPuncttrue
\mciteSetBstMidEndSepPunct{\mcitedefaultmidpunct}
{\mcitedefaultendpunct}{\mcitedefaultseppunct}\relax
\EndOfBibitem
\bibitem[Uiberacker \emph{et~al.}(2007)Uiberacker, Uphues, Schultze, Verhoef,
  Yakovlev, Kling, Rauschenberger, Kabachnik, Schr\"oder, Lezius, Kompa,
  Muller, Vrakking, Hendel, Kleineberg, Heinzmann, Drescher, and
  Krausz]{UibNat07}
M.~Uiberacker, T.~Uphues, M.~Schultze, A.~J. Verhoef, V.~Yakovlev, M.~F. Kling,
  J.~Rauschenberger, N.~M. Kabachnik, H.~Schr\"oder, M.~Lezius, K.~L. Kompa,
  H.-G. Muller, M.~J.~J. Vrakking, S.~Hendel, U.~Kleineberg, U.~Heinzmann,
  M.~Drescher and F.~Krausz, \emph{Nature}, 2007, \textbf{446}, 627\relax
\mciteBstWouldAddEndPuncttrue
\mciteSetBstMidEndSepPunct{\mcitedefaultmidpunct}
{\mcitedefaultendpunct}{\mcitedefaultseppunct}\relax
\EndOfBibitem
\bibitem[Goulielmakis \emph{et~al.}(2010)Goulielmakis, Loh, Wirth, Santra,
  Rohringer, Yakovlev, Zherebtsov, Pfeifer, Azzeer, Kling, Leone, and
  Krausz]{GouNat10}
E.~Goulielmakis, Z.-H. Loh, A.~Wirth, R.~Santra, N.~Rohringer, V.~S. Yakovlev,
  S.~Zherebtsov, T.~Pfeifer, A.~M. Azzeer, M.~F. Kling, S.~R. Leone and
  F.~Krausz, \emph{Nature}, 2010, \textbf{466}, 739\relax
\mciteBstWouldAddEndPuncttrue
\mciteSetBstMidEndSepPunct{\mcitedefaultmidpunct}
{\mcitedefaultendpunct}{\mcitedefaultseppunct}\relax
\EndOfBibitem
\bibitem[Kling \emph{et~al.}(2006)Kling, Siedschlag, Verhoef, Khan, Schultze,
  Uphues, Ni, Uiberacker, Krausz, and Vrakking]{KliSci06}
M.~F. Kling, C.~Siedschlag, A.~J. Verhoef, J.~I. Khan, M.~Schultze, T.~Uphues,
  Y.~Ni, M.~Uiberacker, M.~D.~F. Krausz and M.~J.~J. Vrakking, \emph{Science},
  2006, \textbf{312}, 246\relax
\mciteBstWouldAddEndPuncttrue
\mciteSetBstMidEndSepPunct{\mcitedefaultmidpunct}
{\mcitedefaultendpunct}{\mcitedefaultseppunct}\relax
\EndOfBibitem
\bibitem[Wollenhaupt \emph{et~al.}(2006)Wollenhaupt, Pr\"akelt, Sarpe-Tudoran,
  Liese, and Baumert]{WolAPB06}
M.~Wollenhaupt, A.~Pr\"akelt, C.~Sarpe-Tudoran, D.~Liese and T.~Baumert,
  \emph{Appl. Phys. B}, 2006, \textbf{82}, 183\relax
\mciteBstWouldAddEndPuncttrue
\mciteSetBstMidEndSepPunct{\mcitedefaultmidpunct}
{\mcitedefaultendpunct}{\mcitedefaultseppunct}\relax
\EndOfBibitem
\bibitem[Destefani \emph{et~al.}(2010)Destefani, McDonald, Sukiasyan, and
  Brabec]{DesPRB10}
C.~F. Destefani, C.~McDonald, S.~Sukiasyan and T.~Brabec, \emph{Phys. Rev. B},
  2010, \textbf{81}, 045314\relax
\mciteBstWouldAddEndPuncttrue
\mciteSetBstMidEndSepPunct{\mcitedefaultmidpunct}
{\mcitedefaultendpunct}{\mcitedefaultseppunct}\relax
\EndOfBibitem
\bibitem[Skopalov\'a \emph{et~al.}(2010)Skopalov\'a, El-Taha, Za\"ir,
  Hohenberger, Springate, Tisch, Smith, and Marangos]{SkoPRL10}
E.~Skopalov\'a, Y.~C. El-Taha, A.~Za\"ir, M.~Hohenberger, E.~Springate,
  J.~W.~G. Tisch, R.~A. Smith and J.~P. Marangos, \emph{Phys. Rev. Lett.},
  2010, \textbf{104}, 203401\relax
\mciteBstWouldAddEndPuncttrue
\mciteSetBstMidEndSepPunct{\mcitedefaultmidpunct}
{\mcitedefaultendpunct}{\mcitedefaultseppunct}\relax
\EndOfBibitem
\bibitem[Mathur and Rajgara(2010)]{MatJCP10}
D.~Mathur and F.~A. Rajgara, \emph{J. Chem. Phys.}, 2010, \textbf{133},
  061101\relax
\mciteBstWouldAddEndPuncttrue
\mciteSetBstMidEndSepPunct{\mcitedefaultmidpunct}
{\mcitedefaultendpunct}{\mcitedefaultseppunct}\relax
\EndOfBibitem
\bibitem[Nguyen \emph{et~al.}(2004)Nguyen, Bandrauk, and Ullrich]{NguPRA04}
H.~S. Nguyen, A.~D. Bandrauk and C.~A. Ullrich, \emph{Phys. Rev. A}, 2004,
  \textbf{69}, 063415\relax
\mciteBstWouldAddEndPuncttrue
\mciteSetBstMidEndSepPunct{\mcitedefaultmidpunct}
{\mcitedefaultendpunct}{\mcitedefaultseppunct}\relax
\EndOfBibitem
\bibitem[Fennel \emph{et~al.}(2010)Fennel, Meiwes-Broer, Tiggesbäumker,
  Reinhard, Dinh, and Suraud]{FenRMP10}
T.~Fennel, K.-H. Meiwes-Broer, J.~Tiggesbäumker, P.-G. Reinhard, P.~M. Dinh and
  E.~Suraud, \emph{Rev. Mod. Phys.}, 2010, \textbf{82}, 1793\relax
\mciteBstWouldAddEndPuncttrue
\mciteSetBstMidEndSepPunct{\mcitedefaultmidpunct}
{\mcitedefaultendpunct}{\mcitedefaultseppunct}\relax
\EndOfBibitem
\bibitem[de~Heer \emph{et~al.}(1987)de~Heer, Selby, Kresin, Masui, Vollmer,
  Chatelain, and Knight]{HeePRL87}
W.~A. de~Heer, K.~Selby, V.~Kresin, J.~Masui, M.~Vollmer, A.~Chatelain and
  W.~D. Knight, \emph{Phys. Rev. Lett.}, 1987, \textbf{59}, 1805\relax
\mciteBstWouldAddEndPuncttrue
\mciteSetBstMidEndSepPunct{\mcitedefaultmidpunct}
{\mcitedefaultendpunct}{\mcitedefaultseppunct}\relax
\EndOfBibitem
\bibitem[Yabana and Bertsch(1996)]{Yab96}
K.~Yabana and G.~F. Bertsch, \emph{Phys. Rev. B}, 1996, \textbf{54}, 4484\relax
\mciteBstWouldAddEndPuncttrue
\mciteSetBstMidEndSepPunct{\mcitedefaultmidpunct}
{\mcitedefaultendpunct}{\mcitedefaultseppunct}\relax
\EndOfBibitem
\bibitem[Tiggesb\"aumker \emph{et~al.}(1996)Tiggesb\"aumker, K\"oller, and
  Meiwes-Broer]{TigCPL96}
J.~Tiggesb\"aumker, L.~K\"oller and K.-H. Meiwes-Broer, \emph{Chem. Phys.
  Lett.}, 1996, \textbf{260}, 428--432\relax
\mciteBstWouldAddEndPuncttrue
\mciteSetBstMidEndSepPunct{\mcitedefaultmidpunct}
{\mcitedefaultendpunct}{\mcitedefaultseppunct}\relax
\EndOfBibitem
\bibitem[Schmidt and Haberland(1999)]{SchEPJD99b}
M.~Schmidt and H.~Haberland, \emph{Eur. Phys. J. D}, 1999, \textbf{6},
  109--118\relax
\mciteBstWouldAddEndPuncttrue
\mciteSetBstMidEndSepPunct{\mcitedefaultmidpunct}
{\mcitedefaultendpunct}{\mcitedefaultseppunct}\relax
\EndOfBibitem
\bibitem[Mie(1908)]{Mie08}
G.~Mie, \emph{Ann. Phys. (Leipzig)}, 1908, \textbf{25}, 377\relax
\mciteBstWouldAddEndPuncttrue
\mciteSetBstMidEndSepPunct{\mcitedefaultmidpunct}
{\mcitedefaultendpunct}{\mcitedefaultseppunct}\relax
\EndOfBibitem
\bibitem[Brabec and Krausz(2000)]{BraRMP00}
T.~Brabec and F.~Krausz, \emph{Rev. Mod. Phys.}, 2000, \textbf{72},
  545--592\relax
\mciteBstWouldAddEndPuncttrue
\mciteSetBstMidEndSepPunct{\mcitedefaultmidpunct}
{\mcitedefaultendpunct}{\mcitedefaultseppunct}\relax
\EndOfBibitem
\bibitem[de~Heer(1993)]{Hee93}
W.~A. de~Heer, \emph{Rev. Mod. Phys.}, 1993, \textbf{65}, 611\relax
\mciteBstWouldAddEndPuncttrue
\mciteSetBstMidEndSepPunct{\mcitedefaultmidpunct}
{\mcitedefaultendpunct}{\mcitedefaultseppunct}\relax
\EndOfBibitem
\bibitem[Calvayrac \emph{et~al.}(2000)Calvayrac, Reinhard, Suraud, and
  Ullrich]{Cal00}
F.~Calvayrac, P.-G. Reinhard, E.~Suraud and C.~A. Ullrich, \emph{Phys. Rep.},
  2000, \textbf{337}, 493\relax
\mciteBstWouldAddEndPuncttrue
\mciteSetBstMidEndSepPunct{\mcitedefaultmidpunct}
{\mcitedefaultendpunct}{\mcitedefaultseppunct}\relax
\EndOfBibitem
\bibitem[Tiggesb\"aumker \emph{et~al.}(1993)Tiggesb\"aumker, K\"oller,
  Meiwes-Broer, and Liebsch]{TigPRA93}
J.~Tiggesb\"aumker, L.~K\"oller, K.-H. Meiwes-Broer and A.~Liebsch, \emph{Phys.
  Rev. A}, 1993, \textbf{48}, R1749\relax
\mciteBstWouldAddEndPuncttrue
\mciteSetBstMidEndSepPunct{\mcitedefaultmidpunct}
{\mcitedefaultendpunct}{\mcitedefaultseppunct}\relax
\EndOfBibitem
\bibitem[Scharte \emph{et~al.}(2001)Scharte, Porath, Ohms, Aeschlimann, Krenn,
  Ditlbacher, Aussenegg, and Liebsch]{SchAPB01}
M.~Scharte, R.~Porath, T.~Ohms, M.~Aeschlimann, J.~R. Krenn, H.~Ditlbacher,
  F.~R. Aussenegg and A.~Liebsch, \emph{Appl. Phys. B}, 2001, \textbf{73},
  305\relax
\mciteBstWouldAddEndPuncttrue
\mciteSetBstMidEndSepPunct{\mcitedefaultmidpunct}
{\mcitedefaultendpunct}{\mcitedefaultseppunct}\relax
\EndOfBibitem
\bibitem[Saalmann \emph{et~al.}(2006)Saalmann, Siedschlag, and Rost]{SaaJPB06}
U.~Saalmann, C.~Siedschlag and J.~M. Rost, \emph{J. Phys. B}, 2006,
  \textbf{39}, R39\relax
\mciteBstWouldAddEndPuncttrue
\mciteSetBstMidEndSepPunct{\mcitedefaultmidpunct}
{\mcitedefaultendpunct}{\mcitedefaultseppunct}\relax
\EndOfBibitem
\bibitem[D\"oppner \emph{et~al.}(2005)D\"oppner, Fennel, Diederich,
  Tiggesb\"aumker, and Meiwes-Broer]{DoePRL05}
T.~D\"oppner, T.~Fennel, T.~Diederich, J.~Tiggesb\"aumker and K.-H.
  Meiwes-Broer, \emph{Phys. Rev. Lett.}, 2005, \textbf{94}, 013401\relax
\mciteBstWouldAddEndPuncttrue
\mciteSetBstMidEndSepPunct{\mcitedefaultmidpunct}
{\mcitedefaultendpunct}{\mcitedefaultseppunct}\relax
\EndOfBibitem
\bibitem[D\"oppner \emph{et~al.}(2006)D\"oppner, Fennel, Radcliffe,
  Tiggesb\"aumker, and Meiwes-Broer]{DoePRA06}
T.~D\"oppner, T.~Fennel, P.~Radcliffe, J.~Tiggesb\"aumker and K.-H.
  Meiwes-Broer, \emph{Phys. Rev. A}, 2006, \textbf{73}, 031202(R)\relax
\mciteBstWouldAddEndPuncttrue
\mciteSetBstMidEndSepPunct{\mcitedefaultmidpunct}
{\mcitedefaultendpunct}{\mcitedefaultseppunct}\relax
\EndOfBibitem
\bibitem[Brumer and Shapiro(1986)]{BruCPL86}
P.~Brumer and M.~Shapiro, \emph{Chem. Phys. Lett.}, 1986, \textbf{126},
  541\relax
\mciteBstWouldAddEndPuncttrue
\mciteSetBstMidEndSepPunct{\mcitedefaultmidpunct}
{\mcitedefaultendpunct}{\mcitedefaultseppunct}\relax
\EndOfBibitem
\bibitem[Bouchene \emph{et~al.}(1998)Bouchene, Blanchet, Nicole, Melikechi,
  Girard, Ruppe, Rutz, Schreiber, and W\"oste]{BouEPJD98}
M.~Bouchene, V.~Blanchet, C.~Nicole, N.~Melikechi, B.~Girard, H.~Ruppe,
  S.~Rutz, E.~Schreiber and L.~W\"oste, \emph{Eur. Phys. J. D}, 1998,
  \textbf{2}, 131\relax
\mciteBstWouldAddEndPuncttrue
\mciteSetBstMidEndSepPunct{\mcitedefaultmidpunct}
{\mcitedefaultendpunct}{\mcitedefaultseppunct}\relax
\EndOfBibitem
\bibitem[Blanchet \emph{et~al.}(1997)Blanchet, Nicole, Bouchene, and
  Girard]{BlaPRL97}
V.~Blanchet, C.~Nicole, M.-A. Bouchene and B.~Girard, \emph{Phys. Rev. Lett.},
  1997, \textbf{78}, 2716\relax
\mciteBstWouldAddEndPuncttrue
\mciteSetBstMidEndSepPunct{\mcitedefaultmidpunct}
{\mcitedefaultendpunct}{\mcitedefaultseppunct}\relax
\EndOfBibitem
\bibitem[Scherer \emph{et~al.}(1991)Scherer, Carlson, Matro, Du, Ruggiero,
  Romero-Rochin, Cina, Fleming, and Rice]{SchJCP91}
N.~F. Scherer, R.~J. Carlson, A.~Matro, M.~Du, A.~J. Ruggiero,
  V.~Romero-Rochin, J.~A. Cina, G.~R. Fleming and S.~A. Rice, \emph{J. Chem.
  Phys.}, 1991, \textbf{95}, 1487\relax
\mciteBstWouldAddEndPuncttrue
\mciteSetBstMidEndSepPunct{\mcitedefaultmidpunct}
{\mcitedefaultendpunct}{\mcitedefaultseppunct}\relax
\EndOfBibitem
\bibitem[Katsuki \emph{et~al.}(2009)Katsuki, Chiba, Meier, Girard, and
  Ohmori]{KatPRL09}
H.~Katsuki, H.~Chiba, C.~Meier, B.~Girard and K.~Ohmori, \emph{Phys. Rev.
  Lett.}, 2009, \textbf{102}, 103602\relax
\mciteBstWouldAddEndPuncttrue
\mciteSetBstMidEndSepPunct{\mcitedefaultmidpunct}
{\mcitedefaultendpunct}{\mcitedefaultseppunct}\relax
\EndOfBibitem
\bibitem[Fennel \emph{et~al.}(2004)Fennel, Bertsch, and
  Meiwes-Broer]{FenEPJD04}
T.~Fennel, G.~F. Bertsch and K.-H. Meiwes-Broer, \emph{Eur. Phys. J. D}, 2004,
  \textbf{29}, 367\relax
\mciteBstWouldAddEndPuncttrue
\mciteSetBstMidEndSepPunct{\mcitedefaultmidpunct}
{\mcitedefaultendpunct}{\mcitedefaultseppunct}\relax
\EndOfBibitem
\bibitem[Fennel \emph{et~al.}(2007)Fennel, D\"oppner, Passig, Schaal,
  Tiggesb\"aumker, and Meiwes-Broer]{FenPRL07a}
T.~Fennel, T.~D\"oppner, J.~Passig, C.~Schaal, J.~Tiggesb\"aumker and K.-H.
  Meiwes-Broer, \emph{Phys. Rev. Lett.}, 2007, \textbf{98}, 143401\relax
\mciteBstWouldAddEndPuncttrue
\mciteSetBstMidEndSepPunct{\mcitedefaultmidpunct}
{\mcitedefaultendpunct}{\mcitedefaultseppunct}\relax
\EndOfBibitem
\bibitem[Bertsch and {Das Gupta}(1988)]{BerPR88}
G.~F. Bertsch and S.~{Das Gupta}, \emph{Phys. Rep.}, 1988, \textbf{160},
  189\relax
\mciteBstWouldAddEndPuncttrue
\mciteSetBstMidEndSepPunct{\mcitedefaultmidpunct}
{\mcitedefaultendpunct}{\mcitedefaultseppunct}\relax
\EndOfBibitem
\bibitem[Domps \emph{et~al.}(1997)Domps, L'Eplattenier, Reinhard, and
  Suraud]{DomAP97a}
A.~Domps, P.~L'Eplattenier, P.-G. Reinhard and E.~Suraud, \emph{Ann. Phys.},
  1997, \textbf{6}, 455\relax
\mciteBstWouldAddEndPuncttrue
\mciteSetBstMidEndSepPunct{\mcitedefaultmidpunct}
{\mcitedefaultendpunct}{\mcitedefaultseppunct}\relax
\EndOfBibitem
\bibitem[Plagne \emph{et~al.}(2000)Plagne, Daligault, Yabana, Tazawa, Abe, and
  Guet]{PlaPRA00}
L.~Plagne, J.~Daligault, K.~Yabana, T.~Tazawa, Y.~Abe and C.~Guet, \emph{Phys.
  Rev. A}, 2000, \textbf{61}, 033201\relax
\mciteBstWouldAddEndPuncttrue
\mciteSetBstMidEndSepPunct{\mcitedefaultmidpunct}
{\mcitedefaultendpunct}{\mcitedefaultseppunct}\relax
\EndOfBibitem
\bibitem[Fennel and K\"ohn({2008})]{FenLNP08_}
T.~Fennel and J.~K\"ohn, in {\it Computational Many-Particle Physics}, edited
  by H. Fehske, R. Schneider, and A. Weisse, Lecture notes in physics {\bf
  739}, (Springer Berlin), {2008}, pp. {255--273}\relax
\mciteBstWouldAddEndPuncttrue
\mciteSetBstMidEndSepPunct{\mcitedefaultmidpunct}
{\mcitedefaultendpunct}{\mcitedefaultseppunct}\relax
\EndOfBibitem
\bibitem[Gunnarsson and Lundquist(1976)]{GunPRB76}
O.~Gunnarsson and B.~Lundquist, \emph{Phys. Rev. B}, 1976, \textbf{13},
  4274\relax
\mciteBstWouldAddEndPuncttrue
\mciteSetBstMidEndSepPunct{\mcitedefaultmidpunct}
{\mcitedefaultendpunct}{\mcitedefaultseppunct}\relax
\EndOfBibitem
\bibitem[K\"ohn \emph{et~al.}(2008)K\"ohn, Redmer, Meiwes-Broer, and
  Fennel]{KoePRA08}
J.~K\"ohn, R.~Redmer, K.-H. Meiwes-Broer and T.~Fennel, \emph{Phys. Rev. A},
  2008, \textbf{77}, 033202\relax
\mciteBstWouldAddEndPuncttrue
\mciteSetBstMidEndSepPunct{\mcitedefaultmidpunct}
{\mcitedefaultendpunct}{\mcitedefaultseppunct}\relax
\EndOfBibitem
\end{mcitethebibliography}
\providecommand*{\mcitethebibliography}{\thebibliography}
\csname @ifundefined\endcsname{endmcitethebibliography}
{\let\endmcitethebibliography\endthebibliography}{}

}

\end{document}